\documentclass[]{spie}  

\expandafter\let\csname equation*\endcsname\relax
\expandafter\let\csname endequation*\endcsname\relax
\usepackage[T1]{fontenc}
\usepackage[utf8]{inputenc}
\usepackage{amsmath,amsfonts,amssymb}
\usepackage{braket}
\usepackage{graphicx}
\usepackage{siunitx}
\usepackage[colorlinks=true, allcolors=blue]{hyperref}

\title{Spatial Multiplexing in a Cavity-Enhanced Quantum Memory}

\author[a]{Paul D. Kunz}
\author[a]{Siddhartha Santra}
\author[a]{David H. Meyer}
\author[b]{Zachary A. Castillo}
\author[a]{Vladimir S. Malinovsky}
\author[a]{Kevin C. Cox}
\affil[a]{U.S. Army Research Laboratory, Adelphi, MD 20783, USA}
\affil[b]{Physics Department, University of Maryland, College Park, MD 20742, USA}

\authorinfo{Further author information: (Send correspondence to P.D.K.)\\P.D.K.: E-mail: paul.d.kunz.civ@mail.mil, Telephone: 1 301 394 1980\\}

\begin{document} 
\maketitle

\begin{abstract}
We present a multiplexed quantum repeater protocol based on an ensemble of laser-cooled and trapped rubidium atoms inside an optical ring cavity.  We have already demonstrated strong collective coupling in such a system and have constructed a multiplexing apparatus based on a two-dimensional acousto-optical deflector.  Here, we show how this system could enable a multiplexed quantum repeater using collective excitations with non-trivial spatial phase profiles (spinwaves). Calculated entanglement generation rates over long distances reveal that such a multiplexed ensemble-cavity platform is a promising route towards long distance quantum entanglement and networking.
\end{abstract}


\section{Introduction}
\label{sec:intro}  

Quantum entanglement is the fundamental resource underpinning what is being dubbed the ``Second Quantum Revolution,'' as it enables advances such as sub-shot noise level sensing and metrology, ultra-secure communications, and efficient quantum computation. One of the key aspects of entanglement is the non-local nature of the correlations involved, and researchers have made steady progress creating and manipulating these long-range correlations. A quantum repeater is a device that would facilitate ultra-long range entanglement, but unfortunately its realization has proven elusive despite many subordinate tasks being successfully demonstrated. For example, quantum memories, entanglement swapping, teleportation, and entanglement purification, have been established in a variety of platforms, including neutral ensembles \cite{duan_long-distance_2001,sangouard_quantum_2011}, individual trapped ions \cite{blinov_observation_2004} and atoms \cite{reiserer_cavity-based_2015}, solid state color centers \cite{bernien_heralded_2013}, and rare-earth-doped solids \cite{usmani_mapping_2010,zhou_quantum_2015}. Taking maximum advantage of the inherent strengths of a given platform is critical to realizing such a challenging task as long-distance entanglement distribution. Detailed models have underscored the critical figures of merit that must be addressed: quantum efficiency \cite{gorshkov_universal_2007}, multimode capacity\cite{collins_multiplexed_2007}, and memory lifetime \cite{dudin_light_2013}. 

Quantum efficiency is a broad term characterizing the success probability of each attempt of the protocol and is often separated according to the various constituent components, for instance: 1) inherent memory efficiency of light-matter interface or the probability to create the desired photon from the memory qubit, 2) channel efficiency i.e. fiber or optical loss, and 3) detector efficiency. 
Put simply, efficient interaction between light and matter relies on having lots of light and/or lots of matter, or in other words large field intensity and/or large optical depth (OD). Briefly surveying some long-distance matter-qubit teleportation results reveals the importance of quantum efficiency. In 2009 the quantum state of a trapped ion was teleported to another ion located in a separate chamber one meter away \cite{olmschenk_quantum_2009}, but the efficiency ($10^{-8}$) was severely limited by the light-matter interface. An optical cavity can enhance the field intensity and effective optical depth (via the multipass effect), and this approach led to an increase in teleportation efficiency by five orders of magnitude (to $10^{-3}$), as demonstrated using neutral atom qubits separated by 21 meters \cite{ritter_elementary_2012, nolleke_efficient_2013}. This cavity enhancement cannot be directly transferred to trapped ion platforms because of the ions' extreme sensitivity to charges that accumulate on the nearby cavity mirror surfaces. Alternatively, another free-space approach of using many atoms to increase efficiency demonstrated successful teleportation between neutral ensembles separated by 150 meters of fiber with an efficiency of approximately $10^{-4}$ \cite{bao_efficient_2012}. This brings home the advantages afforded by optical cavities \emph{and} large ensembles of atoms.

The approach of using neutral ensembles has been widely studied due in large part to the proposal by Duan, Lukin, Cirac, and Zoller, known as the DLCZ protocol \cite{duan_long-distance_2001}, which described a method to realize a repeater using relatively modest experimental ingredients. As far as we are aware, the experiment closest to demonstrating a fully functional quantum repeater was that of Kimble's group, essentially attempting this protocol\cite{laurat_towards_2007, choi_mapping_2008}. Unfortunately they were not able to accumulate the statistics to prove entanglement distribution with their repeater, primarily due to the compromise of low entanglement generation probability required to keep errors low (fidelity high). This compromise arises in the first step of the protocol because, for a probability, $p$, to generate a photon that heralds the creation of a spinwave memory (via a Raman scattering process), there is a probability, $p^2$, to generate two excitations leading to an error. The strength of the DLCZ method comes in the second step of the protocol, known as entanglement swapping, where collective excitations (referred to here as "spinwaves") from two neighboring nodes are read out converting the matter excitation back into a photon, which is then interfered on a beam splitter. A successful detector click projects the remote memory nodes into an entangled state, hence the swapping operation doubles the entanglement length. The DLCZ readout or swapping process is efficient because the collective enhancement of the atoms constructively interfering produces a highly directional Readout photon in a known spatial mode, defined by the phase-matching condition (the same physical process as four-wave mixing).  Progress has continued subsequent to this initial attempt at a repeater, notably in the areas of long memory lifetimes \cite{dudin_light_2013,bao_efficient_2012}, various multiplexing schemes \cite{chrapkiewicz_high-capacity_2017,pu_experimental_2017,tian_spatial_2017}, as well as improved repeater protocols that overcome various limitations of the DLCZ scheme\cite{sangouard_quantum_2011}.

Multiplexing is a straightforward and powerful method to greatly increase entanglement distribution rates in a quantum network. Atomic ensembles have an inherent ability for high-capacity multiplexing because, like optical depth, memory mode capacity scales with atom number. Schemes for multiplexing along various degrees of freedom, including temporal, spatial, spectral, internal states, or hybrid combinations thereof \cite{yang_multiplexed_2018} have been experimentally investigated. Spatial multiplexing can be particularly powerful because it also reduces the demands on memory lifetime \cite{collins_multiplexed_2007}. Theoretical models have predicted that reasonably sized ensembles of $\mathit{OD}\approx100$ could support more than $10^3$ memory modes \cite{grodecka-grad_high-capacity_2012}, and recent experimental results have already demonstrated mode capacities greater than $100$ \cite{pu_experimental_2017,parniak_wavevector_2017}. While these demonstrations are impressive they each pose their own challenges when incorporating them into a quantum repeater application scheme. One of the primary challenges being the dynamic optical switching and routing required, and the associated photon loss in any switching technology that exists today.

Here we report a new apparatus, based on neutral atomic ensembles coupled to an optical cavity, that simultaneously integrates solutions to the biggest challenges in realizing a functional quantum repeater. Namely these are: 1)  inefficient light-matter interface, 2) low entanglement generation rates, 3)  rapid accumulation of errors, and 4) stringent phase-stability requirements across the network. We summarize the apparatus and multiplexed quantum repeater scheme.  We then calculate the expected rates for generating remote entanglement over distances up to \SI{800}{\kilo\meter}, showing the large improvements afforded by spinwave multiplexing.

\section{Atom-Cavity Apparatus}

 \begin{figure}[htbp]
  \begin{center}
  \includegraphics[width=16cm]{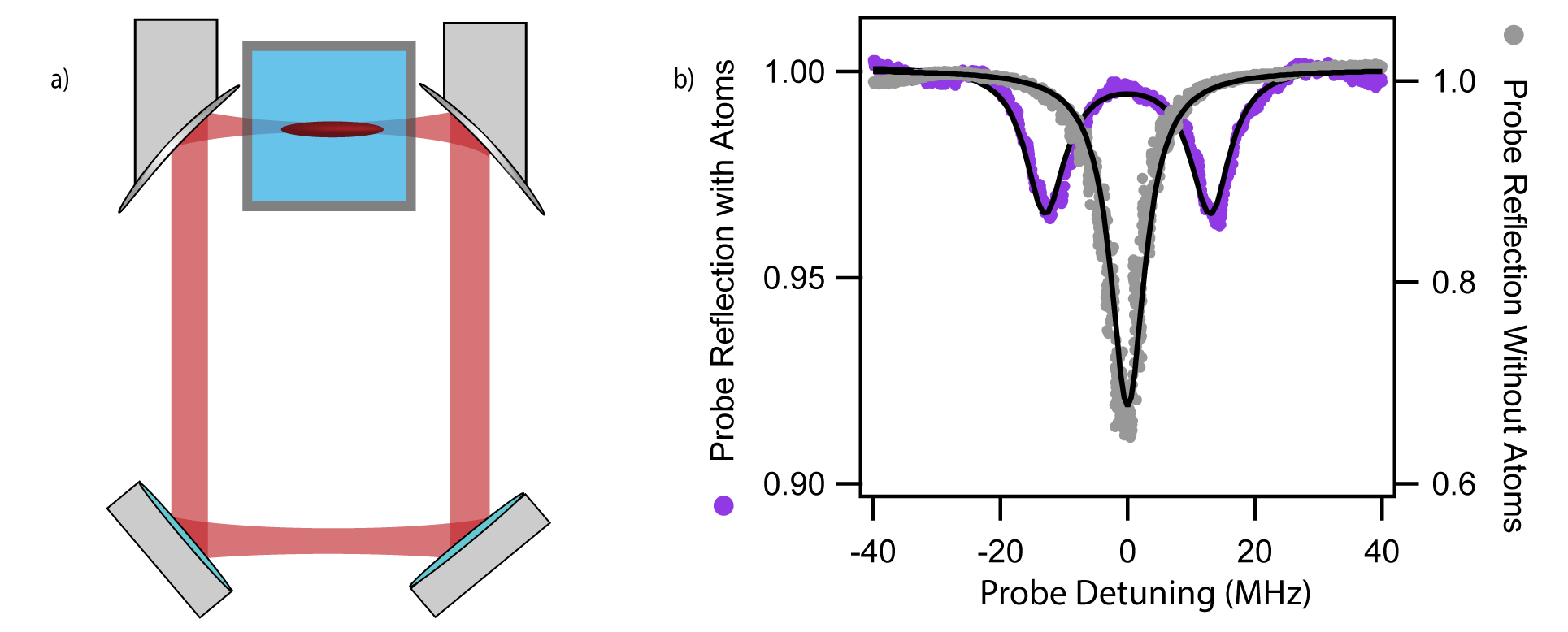}
  \end{center}
  \caption[CavityStuff]
  {\label{fig:CavityStuff}
a) Illustration of our parabolic ring cavity. The cavity mode passes through vacuum cell windows before overlapping with a cold atomic ensemble. The ring geometry leads to greater atom-cavity coupling (smaller waist) while maintaining stability and tunable transverse mode spacing.  b) Fraction of light reflected from bare cavity (gray) and cavity with atoms (purple) as laser frequency is scanned. The vacuum Rabi splitting clearly indicates the strong collective coupling between $10^3$ atoms and the cavity mode. This splitting corresponds to a collective cooperativity of 15, i.e. $94\%$ probability for the photon to be emitted into the cavity mode.}
\end{figure}  

The inherent efficiency of the atomic ensemble light-matter interface is one of its greatest strengths, and stems from the potential for large optical depth ($\mathit{OD}$) \cite{gorshkov_universal_2007}. Classically, $\mathit{OD}$ characterizes the exponential decay of optical transmission, $T=e^{-\mathit{OD}}$, through a medium, and is often written as $\mathit{OD}=\sigma l N/V $. Here $\sigma$ is the interaction cross section, $l$ is the length of the medium, $N$ is the atom number, and $V$ is volume. We can write this in terms of quantum mechanical parameters as,
\begin{align}
\mathit{OD}_{FreeSpace} =N \frac{g^2}{\Gamma}\frac{l}{c},
\label{OD_freespace}
\end{align}
where $g$ is the atom-photon coupling (also called the Jaynes-Cummings parameter), $\Gamma$ is the atomic spontaneous decay rate, and $c$ is the speed of light. Placing the ensemble in a cavity increases the effective $\mathit{OD}$ as the ensemble experiences a multi-pass effect that is proportional to the finesse of the cavity; in other words, we replace the free-space interaction time $l/c$ by the cavity interaction time, a.k.a. cavity lifetime ($1/\kappa$). This cavity-enhanced optical depth is known as the collective cooperativity, $C_N$, 
\begin{align}
\mathit{OD}_{Cavity} =N \frac{g^2}{\Gamma\kappa}\equiv C_N.
\label{OD_cavity}
\end{align}
Cavity-coupled ensembles represent the highest efficiency optical quantum memories yet demonstrated, with $C_N>15$ and intrinsic efficiencies greater than $80\%$ \cite{simon_interfacing_2007,bao_efficient_2012,bimbard_homodyne_2014}.

We have previously reported on our ring cavity design \cite{cox_increased_2018}, and will only review a few salient points here in addition to noting some modifications in our current generation cavity. Traditionally the most popular cavity design in the field of quantum electrodynamics has been the two-mirror Fabry-Perot cavity. The goal typically has been to achieve the single-atom strong coupling regime (defined by $g > \kappa, \Gamma$), where a single atom would exchange energy with a singly excited cavity mode many times before that energy decays to the outside environment. This requires high finesse ($\mathcal{F}$) cavities with small mode volume. Much care is needed to stabilize such cavities against external perturbations, hence it is natural to reduce the complexity and number of optical components as much as possible. The single atom strong coupling regime was achieved many years ago\cite{thompson_observation_1992}, and Fabry-Perot cavity technology has continued to advance - particularly with the advent of fiber-based microcavities \cite{hunger_fiber_2010}. However, the two mirror cavity also brings about several technical compromises. The standing-wave nature of the optical field means that atoms at different locations experience different field amplitudes, hence inhomogeneous light-matter couplings. Additionally, the small radius of curvature of the mirrors means that the surfaces will be close to the matter qubits, which poses a significant challenge particularly for sensitive atomic qubits like ions or Rydberg atoms. Furthermore for single-mode operation, there is a tradeoff between cavity stability range and mode waist size\textemdash to achieve higher field strength (small waist), the stability margin gets sacrificed. Finally, in two-element cavities there is no independent control over transverse mode spacing, and concentric small-waist configurations render these modes degenerate posing a challenge for single-mode operation. A ring cavity design can alleviate these drawbacks \cite{cox_increased_2018}.

The extra degrees of freedom afforded by a travelling-wave ring cavity design means that small mode waists can be achieved while retaining a significantly larger stability margin as compared to Fabry-Perot designs. Since the wave is travelling, this additionally means that the matter qubits experience homogeneous coupling with the field. The mirror surfaces can also be kept further away from the qubits, thus reducing deleterious surface charge effects. Our first generation ring cavity relied on off-axis parabolic focusing mirrors to correct for astigmatism that would otherwise hurt the achievable light-matter coupling strength. This cavity enabled stable single-mode resonance with waists as small as \SI{5}{\micro\meter}. Moreover, we placed this cavity outside the vacuum system (mirror surfaces far from matter qubits) and still observed strong collective coupling ($C_N \approx 15$) despite the limited finesse due to the mode passing through the vacuum windows ($\mathcal{F} \approx 80$). We have since built a second generation atom-cavity system that has higher finesse ($\mathcal{F} \approx 130$) and more atoms ($N\approx 10^5$), albeit with larger cavity mode waist, resulting in collective cooperativities that are routinely $C_N>100$. This system establishes a highly efficient light-matter interface that forms the backbone for our quantum memory experiments.

\section{Angular Multiplexing}

 \begin{figure}[htbp]
  \begin{center}
  \includegraphics[width=16cm]{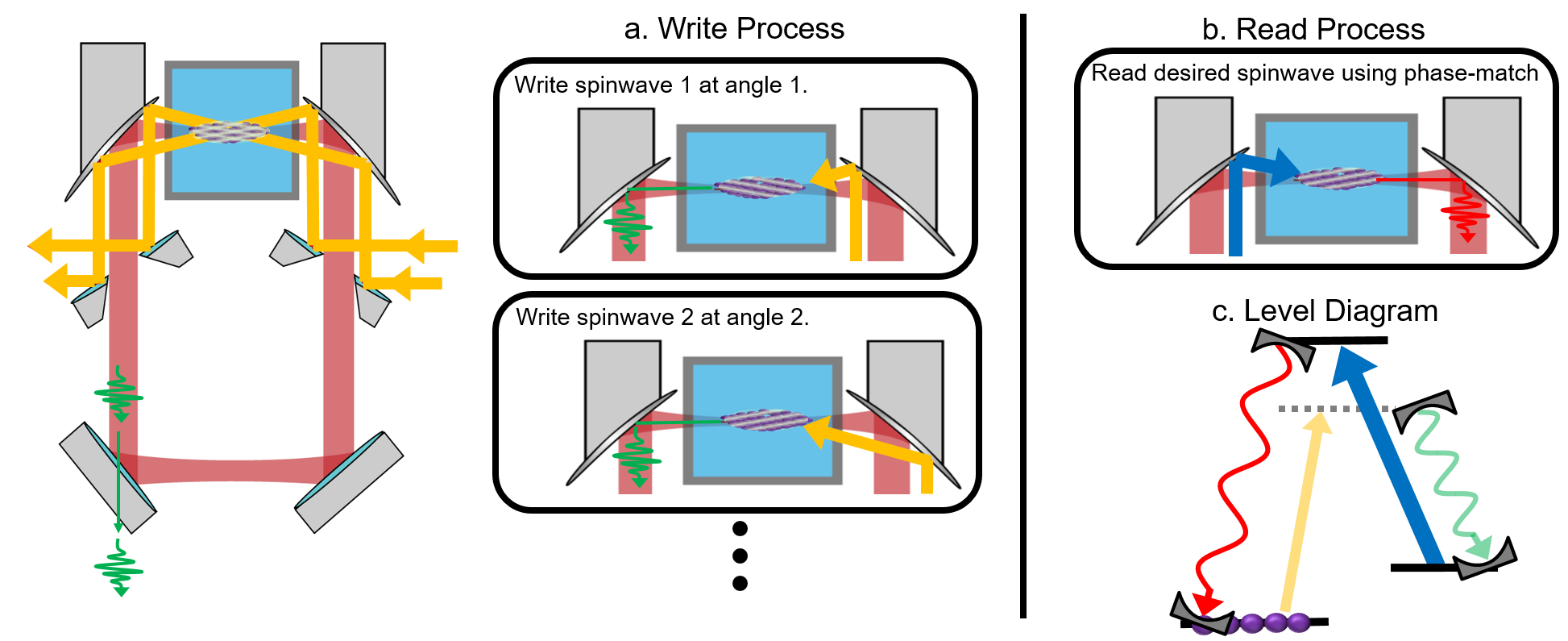}
  \end{center}
  \caption[MultiplexingCartoon]
  {\label{fig:MultiplexingCartoon}
a) Illustration of the multiplexed write process.  The coherent Write beam(s) (red) are imposed at small angles relative to the cavity mode.  A heralding photon (green) is probabilistically emitted into the cavity, leaving behind a collective excitation with a spatial phase grating (purple spinwave). b) Subsequent read process.  The coherent Read laser pulse (blue) is imposed in the opposite direction as the Write laser, leading to a collective enhancement (phase matching) for emission of the Readout photon (red) into the cavity mode.  c) Level diagram.  Cavity modes are tuned to resonance with the Raman transitions associated with the Write and Read processes.}
\end{figure} 

In our system, we trap $10^3$ to $10^5$ $^{87}$Rb atoms in a red-detuned optical trap formed by the cavity mode and laser cool them to approximately \SI{20}{\micro\kelvin}.  The atoms are optically pumped to the lower ground state $\ket{5 S_{1/2}, F=1}$.  The Write step, detuned from the D2 transition, causes cavity-enhanced Raman transitions to the upper ground state ($\ket{5 S_{1/2}, F=2}$), and creates a heralding photon in the cavity mode with probability $p$.  After successfully heralding a spinwave, the subsequent Read step induces the inverse Raman transition, which is collectively enhanced. The Heralding optical cavity mode and Read optical cavity modes are distinct in both directionality and frequency (longitudinal mode number), as each is tuned to be resonant for its respective Raman transition.

The collective spinwave excitations created in the ensemble are such that a single excitation is shared among all the atoms.  The spinwave wavefunction is importantly characterized by a spatial phase $\phi(x) = \Delta \vec{k} \cdot \vec{x}$ where $\Delta \vec{k}$ is the difference in wavevector between the Write laser and cavity mode.  Multiplexed spinwaves are orthogonal if $\phi(x)$ destructively interferes when integrated over the entire atomic cloud. Here we present our proposal for an efficient multiplexed quantum memory node, based on this apparatus, well suited for long distance entanglement distribution.

Significant previous work has been done developing multiplexing schemes based on various degrees of freedom, such as temporal, spatial, or spectral. Subdividing a large cold-atom cloud into smaller sub-ensembles that can be individually addressed using optical switching was first demonstrated in 2009 with twelve resolved micro-ensembles \cite{lan_multiplexed_2009}. A separate group recently expanded this technique demonstrating 225 sub-ensembles \cite{pu_experimental_2017}, and verifying up to 22-partite W-state entanglement \cite{pu_experimental_2018}, but there is drawback in that the OD (and hence intrinsic efficiency) of each memory is much reduced from that of the total ensemble. Another powerful approach uses angularly resolvable spinwaves in a single atomic ensemble. A basic two-mode version of this idea was first demonstrated by Chen et al. \cite{chen_demonstration_2007}, where a single write beam impinges on an atomic ensemble and two separate spatial modes are monitored for heralding photons. This method has been expanded to six modes in a quantum memory that demonstrated a direct increase in the atom-photon entanglement generation rate \cite{tian_spatial_2017}. Finally, observation of quantum correlations in a cold-atom system with 665 resolved angular modes \cite{parniak_wavevector_2017} represents the dramatic potential of this platform, though in this instance the apparatus was not directly suitable as a quantum memory node due to the challenge of mapping the modes to quantum channels (fibers) for long distance entanglement distribution.

In a sense, our scheme can be viewed as the inverse of the angular spinwave methods just mentioned, where a single write beam is sent in and many heralding channels are monitored. Our scheme, as illustrated in Figure \ref{fig:MultiplexingCartoon}, sequentially maps many write beams, each at a distinct angle, to our ensemble and monitors the heralding photons exiting via the cavity mode. These cavity-mode heralding photons are well suited to efficiently couple to a single-mode fiber for distribution and avoid the critically lossy and complex switching components of the previous schemes. That is, rather than having the valuable single photons routed through switches, they instead cleanly exit through the cavity. In our case, the lossy switching component, a two-dimensional acousto-optic deflector (2D AOD), is instead implemented on the classical Write and Read beams where the losses are inconsequential. The phase-matching condition of the read process ensures that the subsequent readout photon also exits through the cavity (see Fig. \ref{fig:MultiplexingCartoon} b), taking advantage of the cavity-enhanced collective emission efficiency. The mode capacity of similar systems has been studied previously, both experimentally \cite{parniak_wavevector_2017} (as mentioned in the previous paragraph) and theoretically \cite{grodecka-grad_high-capacity_2012}, and is known to increase with optical depth and Fresnel number of the atom cloud. The Fresnel number characterizes the spatial geometry of the cloud, and is defined as $F= a^2 / l\lambda$, where $a$ is the transverse radius of the cloud, $l$ is the axial length, and $\lambda$ is the wavelength of the light. Based on these previous studies a conservative estimate of over 100 modes appears readily achievable in our apparatus.

 \begin{figure}[htbp]
  \begin{center}
  \includegraphics[width=8cm]{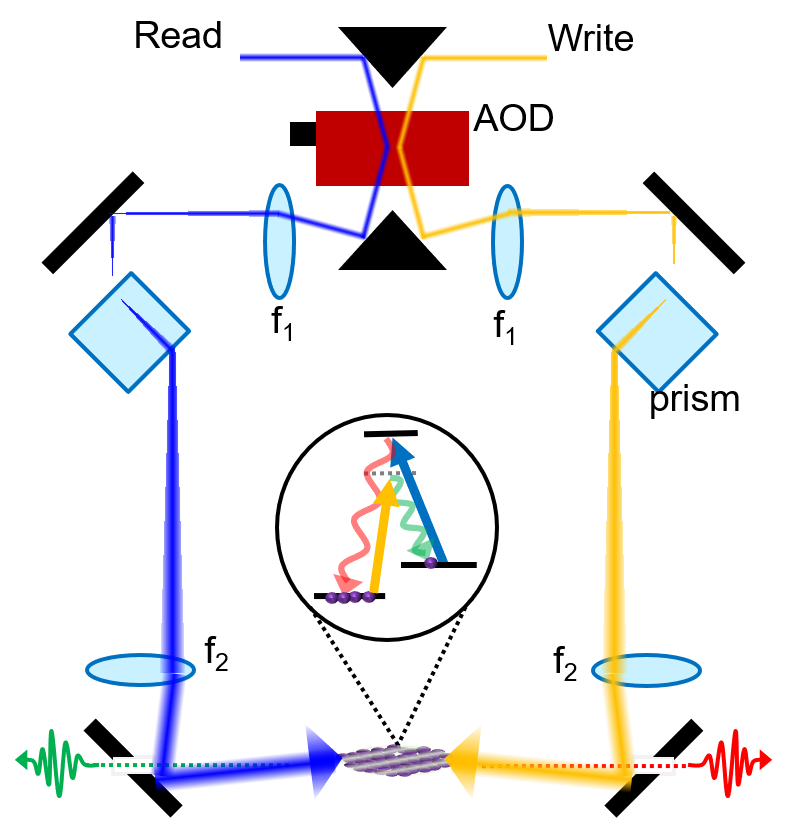}
  \end{center}
  \caption[OptEng]
  {\label{fig:OptEng}
a) Illustration of the Write/Read beams optical system.  The angle of the Read (blue) and Write (yellow) beams are controlled by a two-dimensional acousto-optic deflector (2D-AOD).  The collimated beams at the AOD are relayed to the atoms using a "4f" imaging system with magnification of $f_2/f_1$. The mirrors adjacent the atom cloud have holes in the center for the cavity mode to pass through. The Write/Read beams are displaced from center using prisms in order to avoid the holes.}
\end{figure} 

In order to write and read the desired modes in a precise, controlled, and repeatable way we use an optical system based on a 2D-AOD, diagrammed in Figure \ref{fig:OptEng}. Because the acoustic velocity in the deflector crystal is slow (\SI{0.5}{\milli\meter\per\micro\second}), it is desirable to keep the beam diameter in the crystal small to reduce the time between pulses. At the other end of the system a uniform plane wave evenly impinging on all the atoms is desired in order to maximize the collective interference effect of the four-wave mixing readout process. Therefore we use a "4f" imaging relay system (i.e. the distance from object to image is $2(f_1+f_2)$) with magnification ($f_2/f_1$) to map the angles of the collimated beams at the AOD to larger collimated beams at the atom cloud. An additional acousto-optic modulator (not shown) is placed up stream to compensate for the small differences in frequency between each beam angle that result from the different radio-frequency tones used in the AOD.  

For our first series of experiments we can avoid having to add a lattice trap while still having usable spinwave lifetimes by keeping the spinwave wavelength as long as possible. This is done by using small angles between the Write and Heralding photon modes where the longer spinwave wavelengths reduce sensitivity to atomic motional dephasing. For even longer lifetimes (\SI{>1}{\second}) we will need to confine the atoms with an optical lattice. Our current solution to achieve these small angles, keep large optical access for many modes, and not compromise the cavity mode finesse, was to drill a hole in the center of two mirrors on either side of the atom cloud. The cavity mode passes through the holes, while the various Write and Read beams reflect off the mirror surfaces. To prevent the holes from clipping the write and Read beams we could use large deflection angles in the AOD, but this reduces the usable AOD bandwidth (the usable number of resolvable angles). Instead we place prisms in the 4f imaging system, as shown in figure \ref{fig:OptEng}, which deflect beams around the holes while retaining the full bandwidth of the AOD.

\section{Quantum Repeater}

Building upon the multiplexed and efficient light-matter interface of the previous sections, we now describe a quantum memory node capable of robust two-photon entanglement generation and swapping. 
Such a node would be well suited to perform as a quantum repeater, especially when incorporating the long memory lifetimes already demonstrated in neutral ensembles \cite{dudin_light_2013,bao_efficient_2012}, and their compatibility with telecom wavelengths \cite{radnaev_quantum_2010,farrera_nonclassical_2016}.

\subsection{The Repeater Node}

 \begin{figure}[htbp]
  \begin{center}
  \includegraphics[width=16cm]{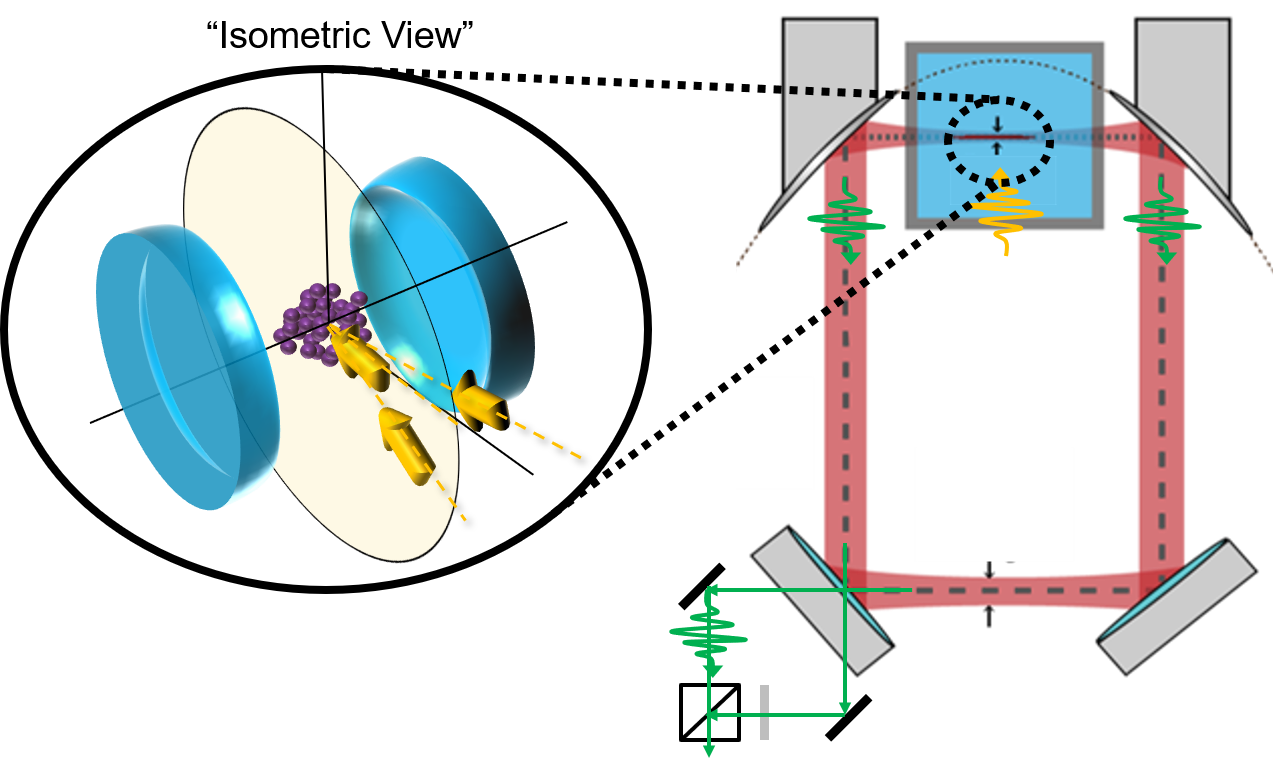}
  \end{center}
  \caption[RepeatNode]
  {\label{fig:RepeatNode}
a) Illustration of our repeater node, with an isometric view showing Write pulses impinging upon the atoms from three different angles. Heralding photons are emitted into the cavity mode (in either direction), whose axis is orthogonal to the plane containing the Write pulses. The opposing right and left cavity modes are then shown to straightforwardly map to a polarization qubit using half-wave retarder and polarizing beam splitter (alternatively they could map to time-bin qubit).}
\end{figure} 

For simplicity, consider a repeater node having the Write beams impinge the atoms from the plane perpendicular to the cavity axis, as shown in the isometric view inset of figure (\ref{fig:RepeatNode}). Note that currently our apparatus is configured for small angles, rather than perpendicular angles, between the Write modes and Heralding mode k-vectors; this is to reduce the spinwaves' sensitivity to atomic motional dephasing as discussed in the previous section, since we have not yet installed an optical lattice trap. In the perpendicular configuration there is symmetry between right- and left-going cavity mode directions for the heralding photons and corresponding spinwaves. For each attempted Write pulse, there is an equal probability to create a right or left spinwave-herald-photon pair, and we sequentially attempt different Write pulse angles in time. 

Each individual Write pulse has a low probability to successfully herald a spinwave memory, but by attempting many modes we increase the overall success rate by a factor equal to the mode capacity, $M$, of our memory. We will need to keep track of which angle corresponds to which time so that we know which mode we can successfully read out for entanglement swapping. This means that there will be synchronized clocks at the memory node and the remote Bell state detector that will indicate which spinwave is successfully heralded. We use lowercase greek letters to denote the successfully heralded spinwave-photon mode pair, with bars above to indicate right-going and no-bars to indicate left-going modes. The right or left going photon momenta can straightforwardly be mapped to a polarization (or time-bin) qubit, $H \text{ or } V$, using a half-wave retarder plate and polarizing beam splitter (PBS) as shown in Figure (\ref{fig:RepeatNode}). Bosonic creation operators for spinwaves and photons are written using conventional notation, $s^{\dagger}$ and $a^{\dagger}$ respectively, as in the review paper of Reference \citenum{sangouard_quantum_2011}. In this way, we can write the state created in repeater node $A$ as,
\begin{align}
\ket{\psi_A}&=\ket{00}+\sqrt{\frac{p}{2}}(s^{\dagger}_{A,\alpha}a^\dagger_{H,\alpha}+s^{\dagger}_{A,\bar{\alpha}}a^\dagger_{V,\bar{\alpha}})\ket{00}+O(p).
\label{ensemble_excite}
\end{align}
Equation \ref{ensemble_excite} expresses the fact that, with probability $p$, one is able to generate the maximally entangled state of a spinwave and photon, $\frac{1}{\sqrt{2}}{(s^{\dagger}_{\alpha}a^\dagger_{H}+s^{\dagger}_{\bar{\alpha}}a^\dagger_{V})}\ket{00}$, where $\alpha$ represents the successful mode out of the total $M$ modes attempted. Excitations of two or more spinwaves are included in the term proportional to $O(p)$, which we assume to be negligibly small.

\subsection{Entanglement Generation Between Remote Nodes}

As DLCZ is one of the most commonly studied protocols, and represents the closest to a successful demonstration of a repeater, we now compare some of the relative merits of our scheme to the DLCZ method.

The DLCZ protocol probabilistically creates remote entanglement between neighboring stations by creating an excitation that is ``shared'' between the two sites.  This entanglement in the ``excitation number'' basis must be created with very low probability to prevent accidental creation of multiple excitations that would drastically lower the entanglement fidelity. The entanglement in the particle-number basis is converted into more readily usable ``Bell-like'' entanglement by doubling the network resources along two parallel channels that are finally read-out in the last step, known as post-selection.

The intrinsically low rate of entanglement generation between remote memory nodes has thus far prevented successful demonstration of a quantum repeater, and is one of the major challenges to not just the DLCZ protocol, but many other repeater schemes in general \cite{laurat_towards_2007}. Multiplexing addresses this challenge head on. Further, our ability to naturally create a Bell-like state between a photon and matter qubit brings additional significant benefits, including: reduction of quantum channel phase sensitivity, reduction of vacuum errors, and elimination of the doubled resources and subsequent post-selection step.

Most importantly, the single-photon interference method for entanglement generation, as used in the DLCZ protocol, requires that the two fiber arm lengths must remain equal to within a fraction of an optical wavelength. This long distance interferometer erases the which-path information of the photon, leading to the shared excitation probability for the remote nodes. Any path length difference introduces the opportunity to learn which node produced the photon and thereby compromises the entangling operation. Chen et al. \cite{chen_fault-tolerant_2007} pointed out that using a two-photon interference method reduces that criteria to just the coherence length of the light, i.e. by roughly seven orders of magnitude. This reduction in sensitivity to quantum channel phase fluctuations is critical for any real-world implementation of a quantum repeater network. In light of this, channel phase sensitivity must be accounted for in order to make fair comparisons between different protocols.

\begin{figure}[htbp]
  \begin{center}
  \includegraphics[width=16cm]{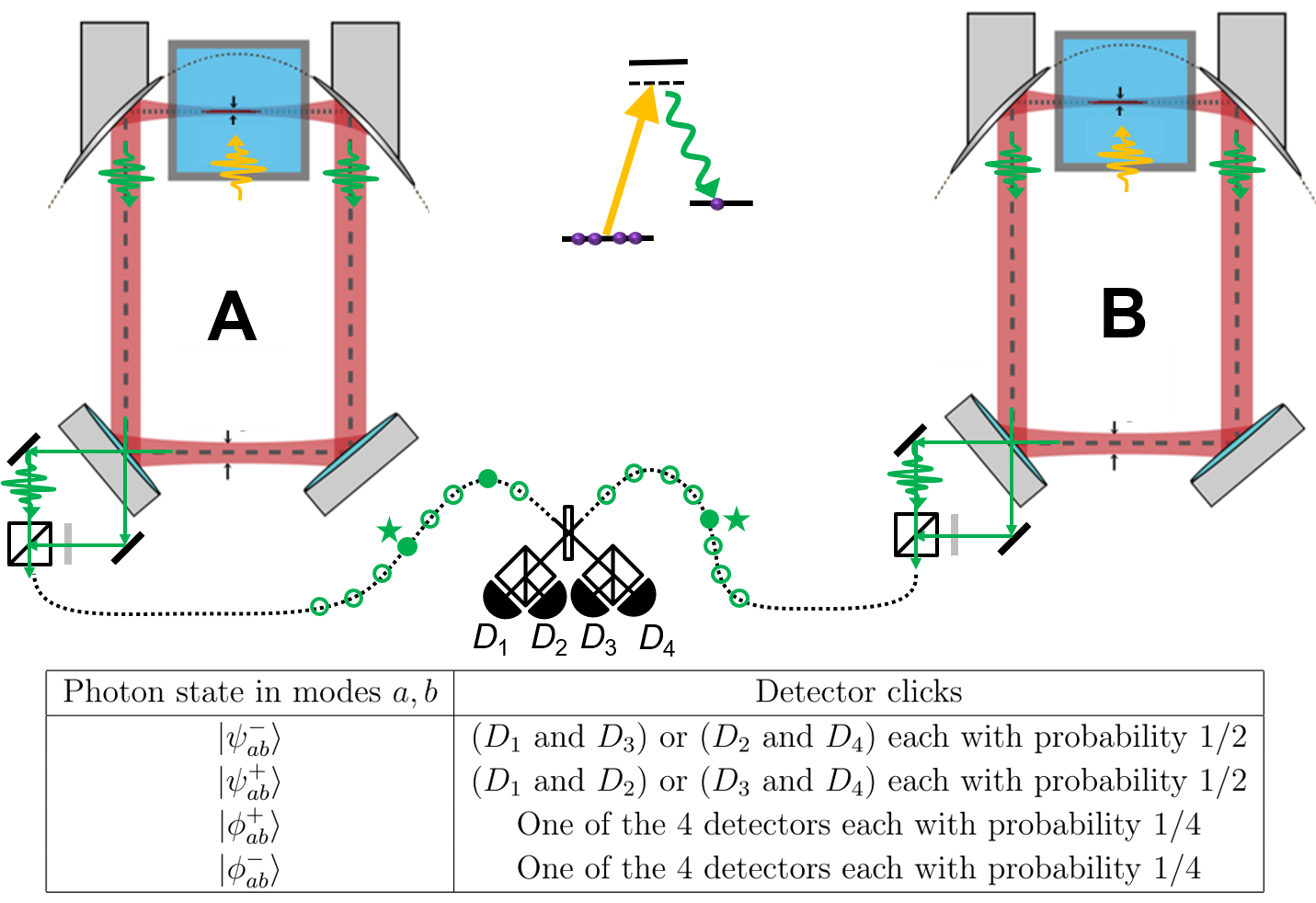}
  \end{center}
  \caption[EntDist]
  {\label{fig:EntDist}
a) Illustration of the entanglement generation process.  Entanglement generation is repeatedly tried using the spatial multiplexing scheme.  Unsuccessful attempts (no heralding photon) are shown as open green circles, and successful attempts are shown as closed circles.  Entanglement is generated when two heralding photons are collected in the same trial, and result in the appropriate pair of detector clicks. The table shows Bell State Measurement (BSM) detector click patterns that correspond to the standard Bell states, as defined in the text.}
\end{figure}

The entanglement generation step for our scheme relies on two-photon interference, as shown in Figure (\ref{fig:EntDist}). Each node, $A$ and $B$, is a copy of the node shown in Figure (\ref{fig:RepeatNode}). The synchronized attempts to Write memory excitations are represented by green circles traveling along the fiber, and solid circles correspond to excitations that generate a heralding photon in the cavity mode (i.e. successfully charged spinwave modes). Of the eight attempts illustrated, only the fifth attempt (highlighted with a star) can create remote entanglement because a heralding photon has been successfully sent from each node. In the event of simultaneous creation of heralding photons, a conventional linear-optical Bell-state measurement (BSM) system at a central station can then project the remote memory nodes into a maximally entangled state, though with only up to $50\%$ efficiency \cite{lutkenhaus_bell_1999}. The table in Figure (\ref{fig:EntDist}) describes the photon detector click patterns that correspond to the standard Bell states $\ket{\psi^\pm_{ab}}=\frac{1}{\sqrt{2}}(\ket{0_{a}1_{b}}\pm\ket{1_{a}0_{b}})$ and $\ket{\phi^\pm_{ab}}=\frac{1}{\sqrt{2}}(\ket{0_{a}0_{b}}\pm\ket{1_{a}1_{b}})$.

The two-photon sector of the joint state of the two ensembles is given by,
\begin{align}
\ket{\psi_A}\ket{\psi_B}|_{ab}=p\frac{(s^{\dagger}_{A,\alpha}a^\dagger_{H}+s^{\dagger}_{A,\bar{\alpha}}a^\dagger_{V})}{\sqrt{2}}\frac{(s^{\dagger}_{B,\beta}b^\dagger_{H}+s^{\dagger}_{B,\bar{\beta}}b^\dagger_{V})}{\sqrt{2}}\ket{0000}.
\label{ABstate2photon}
\end{align}

We will write the state (Eq. \ref{ABstate2photon}) in a form that allows easy analysis of the BSM on the two-photon part of the state. Expanding the operator products in the RHS of (Eq. \ref{ABstate2photon}) we have,
\begin{align}
\ket{\psi_A}\ket{\psi_B}|_{ab}&=\frac{p}{2}(s^\dagger_{A,\alpha}s^\dagger_{B,\beta}a^\dagger_Hb^\dagger_H+s^\dagger_{A,\alpha}s^\dagger_{B,\bar{\beta}}a^\dagger_Hb^\dagger_V+s^\dagger_{A,\bar{\alpha}}s^\dagger_{B,\beta}a^\dagger_Vb^\dagger_H+s^\dagger_{A,\bar{\alpha}}s^\dagger_{B,\bar{\beta}}a^\dagger_Vb^\dagger_V)\ket{0000}\nonumber\\
&=\frac{p}{2}(s^\dagger_{A,\alpha}s^\dagger_{B,\beta}\frac{\ket{\phi^+_{ab}}+\ket{\phi^-_{ab}}}{\sqrt{2}}+s^\dagger_{A,\alpha}s^\dagger_{B,\bar{\beta}}\frac{\ket{\psi^+_{ab}}+\ket{\psi^-_{ab}}}{\sqrt{2}}\nonumber\\
&~~~~~~~~~~~~+s^\dagger_{A,\bar{\alpha}}s^\dagger_{B,\beta}\frac{\ket{\psi^+_{ab}}-\ket{\psi^-_{ab}}}{\sqrt{2}}+s^\dagger_{A,\bar{\alpha}}s^\dagger_{B,\bar{\beta}}\frac{\ket{\phi^+_{ab}}-\ket{\phi^-_{ab}}}{\sqrt{2}})\ket{00}\nonumber\\
&=\frac{p}{2}(\frac{s^\dagger_{A,\alpha}s^\dagger_{B,\bar{\beta}}+s^\dagger_{A,\bar{\alpha}}s^\dagger_{B,\beta}}{\sqrt{2}}\ket{\psi^+_{ab}}+\frac{s^\dagger_{A,\alpha}s^\dagger_{B,\bar{\beta}}-s^\dagger_{A,\bar{\alpha}}s^\dagger_{B,\beta}}{\sqrt{2}}\ket{\psi^-_{ab}}\nonumber\\
&~~~~~~~~~~~~+\frac{s^\dagger_{A,\alpha}s^\dagger_{B,\beta}+s^\dagger_{A,\bar{\alpha}}s^\dagger_{B,\bar{\beta}}}{\sqrt{2}}\ket{\phi^+_{ab}}+\frac{s^\dagger_{A,\alpha}s^\dagger_{B,\beta}-s^\dagger_{A,\bar{\alpha}}s^\dagger_{B,\bar{\beta}}}{\sqrt{2}}\ket{\phi^-_{ab}})\ket{00}\nonumber\\
&=\frac{p}{2}(\ket{\psi^+_{AB}}\ket{\psi^+_{ab}}+\ket{\psi^-_{AB}}\ket{\psi^-_{ab}}+\ket{\phi^+_{AB}}\ket{\phi^+_{ab}}+\ket{\phi^-_{AB}}\ket{\phi^-_{ab}})
\end{align}
Conditioned on appropriate coincident detector clicks one gets $\ket{\psi^+_{AB}}$ or $\ket{\psi^-_{AB}}$ with probability $(p/2)^2$ each. Therefore, the total probability of obtaining an entangled state of the two atomic ensembles is,
\begin{align}
P_0=\frac{1}{2}Mp^2\eta^2_d\eta^2_t
\end{align}
where $M$ is the mode capacity of the quantum memory, $\eta_d$ is the detector efficiency, and $\eta_t=exp(-L_0/2L_{att})$ is the efficiency of the channel transmission. The cycle time of entanglement generation is denoted as $\tau_C$ and is approximately  $(L_0/c)$. $M$ can be limited by either of two different constraints: the number of pulses that can be attempted in the time before the first pulse outcome reaches the BSM, or the number of resolvable spinwave angles that the ensemble is capable of supporting. Hence, we have an upper bound on $M$ of 
\begin{align}
M\leq \textrm{Min}(\tau_c/\tau_w,4\pi/\Omega_{res})
\end{align}

The quantum frequency conversion that is required to achieve the favorable and often quoted fiber attenuation length of $L_{att} = \SI{22}{\kilo\meter}$ is a commonly overlooked source of loss in many repeater protocol rate calculations. This length corresponds to the \SI{0.2}{\decibel\per\kilo\meter} loss rate for \SI{1550}{\nano\meter} wavelength light in standard telecom fiber. Most other wavelengths, especially those of today's best quantum memories, have much more severe losses. For example, the loss rate for a common rubidium wavelength of \SI{780}{\nano\meter} is \SI{3.5}{\decibel\per\kilo\meter} $(L_{att} = \SI{1.2}{\kilo\meter})$, and for ionic barium (a strong interface ion candidate) at \SI{650}{\nano\meter} is \SI{15}{\decibel\per\kilo\meter} $(L_{att} = \SI{0.3}{\kilo\meter})$. Thus it can be favorable to convert the memory photons to the telecom wavelength, as it is possible to preserve the quantum properties through the frequency conversion process. While quantum frequency conversion research is still in early stages, rubidium wavelengths have witnessed significant progress. One demonstration used fourwave mixing in a separate ensemble of rubidium atoms to convert to \SI{1367}{\nano\meter} (corresponding to \SI{0.4}{\decibel\per\kilo\meter}) with $54 \%$ efficiency \cite{radnaev_quantum_2010}. More recently solid-state integrated waveguides have demonstrated conversion to \SI{1550}{\nano\meter}, with overall efficiencies approaching $10\%$ and signal-to-noise ratios $\approx 100$, such that quantum coherence is preserved  \cite{farrera_nonclassical_2016}. Although such conversion efficiency might seem discouraging, these numbers continue to improve. Even with $10 \%$ efficiency, however, it becomes favorable to perform the conversion for fiber lengths greater than \SI{\sim3}{\kilo\meter}.

\subsection{Entanglement swapping}
As discussed above, memory efficiency, or the ability to retrieve a photon from the memory, is one of the most critical figures of merit for any repeater.  This is one of the strengths of repeaters based on ensembles of neutral atoms.  The presence of many atoms give a collective enhancement in the readout process that leads to near unity retrieval efficiency.

On the other hand, a drawback in the scalability of the original DLCZ protocol is the rapid growth of vacuum state errors: errors where a heralding photon is retrieved, but a usable quantum memory is not charged. Due to the reliance on single-photon interference, the vacuum component essentially doubles with every level of entanglement swapping. It was recognized by Jiang et al. \cite{jiang_fast_2007} in 2007 that two-photon interference, as used in our system, mitigates this vacuum error growth keeping it constant regardless of the number of swapping procedures, thus the entanglement distribution rate is significantly increased.

To calculate the probability of successful entanglement swapping in our system, we begin by considering entangled states of two pairs of ensembles, $\ket{\psi^+_{AB}}$ and $\ket{\psi^+_{CD}}$ where,
\begin{align}
\ket{\psi^-_{AB}}&=\frac{s^\dagger_{A,\alpha}s^\dagger_{B,\bar{\beta}}+s^\dagger_{A,\bar{\alpha}}s^\dagger_{B,\beta}}{\sqrt{2}}\ket{00},\nonumber\\
\ket{\psi^-_{CD}}&=\frac{s^\dagger_{C,\alpha}s^\dagger_{D,\bar{\beta}}+s^\dagger_{C,\bar{\alpha}}s^\dagger_{D,\beta}}{\sqrt{2}}\ket{00}.
\end{align}
The read out process for entanglement swapping involves sending a Read laser pulse on ensembles $B$ and $C$ resulting in the emission of a Readout photon. The experimental setup couples the polarization of the Readout photon to the direction of the spin-wave in an opposite manner to that for the Heralding photon. That is, if spin-wave $\alpha$ is associated with the $H$-polarization of the Heralding photon, then upon readout it is associated with the $V$-polarization of the Readout photon. The joint state of ensembles $A,B,C,D$ after a successful readout for $B$ and $C$ just before entanglement swapping is therefore,
\begin{align}
\ket{\psi_{ABCD}}&=\frac{(s^\dagger_{A,\alpha}b^\dagger_H+s^\dagger_{A,\bar{\alpha}}b^\dagger_V)}{\sqrt{2}}\frac{(c^\dagger_Vs^\dagger_{D,\bar{\beta}}+c^\dagger_Hs^\dagger_{D,\beta})}{\sqrt{2}}\ket{0000}\nonumber\\
&=\frac{1}{2}(s^\dagger_{A,\alpha}s^\dagger_{D,\bar{\beta}}b^\dagger_Hc^\dagger_V+s^\dagger_{A,\alpha}s^\dagger_{D,\beta}b^\dagger_Hc^\dagger_H+s^\dagger_{A,\bar{\alpha}}s^\dagger_{D,\bar{\beta}}b^\dagger_Vc^\dagger_V+s^\dagger_{A,\bar{\alpha}}s^\dagger_{D,\beta}b^\dagger_Vc^\dagger_H)\ket{0000}\nonumber\\
&=\frac{1}{2}(s^\dagger_{A,\alpha}s^\dagger_{D,\bar{\beta}}\frac{\ket{\psi^+_{bc}}+\ket{\psi^-_{bc}}}{\sqrt{2}}+s^\dagger_{A,\alpha}s^\dagger_{D,\beta}\frac{\ket{\phi^+_{bc}}+\ket{\phi^-_{bc}}}{\sqrt{2}}\nonumber\\
&~~~~~~~~~~~~~+s^\dagger_{A,\bar{\alpha}}s^\dagger_{D,\bar{\beta}}\frac{\ket{\phi^+_{bc}}-\ket{\phi^-_{bc}}}{\sqrt{2}}+s^\dagger_{A,\bar{\alpha}}s^\dagger_{D,\beta}\frac{\ket{\psi^+_{bc}}-\ket{\psi^-_{bc}}}{\sqrt{2}})\ket{00}\nonumber\\
&=\frac{1}{2}(\frac{s^\dagger_{A,\alpha}s^\dagger_{D,\bar{\beta}}+s^\dagger_{A,\bar{\alpha}}s^\dagger_{D,\beta}}{\sqrt{2}}\ket{\psi^+_{bc}}+\frac{s^\dagger_{A,\alpha}s^\dagger_{D,\bar{\beta}}-s^\dagger_{A,\bar{\alpha}}s^\dagger_{D,\beta}}{\sqrt{2}}\ket{\psi^-_{bc}}\nonumber\\
&~~~~~~~~~~~~~+\frac{s^\dagger_{A,\alpha}s^\dagger_{D,\beta}+s^\dagger_{A,\bar{\alpha}}s^\dagger_{D,\bar{\beta}}}{\sqrt{2}}\ket{\phi^+_{bc}}+\frac{s^\dagger_{A,\alpha}s^\dagger_{D,\beta}-s^\dagger_{A,\bar{\alpha}}s^\dagger_{D,\bar{\beta}}}{\sqrt{2}}\ket{\phi^-_{bc}})\ket{00}\nonumber\\
&=\frac{1}{2}(\ket{\psi^+_{AD}}\ket{\psi^+_{bc}}+\ket{\psi^-_{AD}}\ket{\psi^-_{bc}}+\ket{\phi^+_{AD}}\ket{\phi^+_{bc}}+\ket{\phi^-_{AD}}\ket{\phi^-_{bc}})
\label{swapstate}
\end{align}

Thus, upon a Bell state measurement of the state in Equation \ref{swapstate}, one gets $\ket{\psi^+_{AD}}$ or $\ket{\psi^-_{AD}}$ with probability $1/4$ each. Therefore, the total probability of a successful entanglement swap is,
\begin{align}
P_1=\frac{1}{2}\eta_m^2\eta_d^2
\end{align}

\section{Rate of Entanglement Distribution}

In the absence of multi-photon excitations during the Write step, our protocol has unit fidelity.  Moreover, the remote entangled states $\ket{\psi^{\pm}_{AD}}$ are already in the Bell basis, and no final post-selection is needed, which is necessary to the DLCZ protocol. The rate of entanglement generation in our scheme is therefore,
\begin{align}
T_{\textrm{eg}}=\frac{L_0}{c}\frac{f_0f_1...f_n}{P_0P_1...P_n} \approx \frac{L_0}{c}\frac{(3/2)^{n+1}}{P_0(P_1)^n}
\end{align}
where $n$ is the maximum number of nesting levels, $L_{\textrm{Tot}}=2^nL_0$ is the total distance between furthest nodes, and the numerical factors $f_i$ account for the possible ways that entanglement can be generated at each level.  These numerical factors are approximately $3/2$ at all levels.

 \begin{figure}[htbp]
  \begin{center}
  \includegraphics[width=14cm]{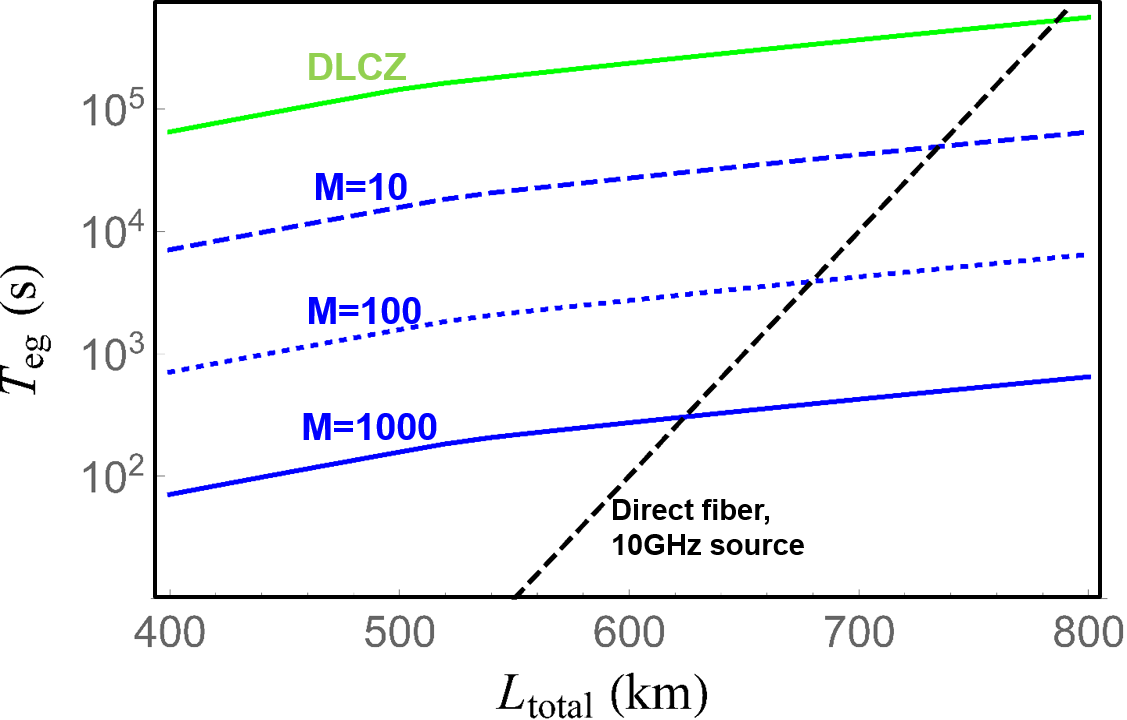}
  \end{center}
  \caption{Time to deterministically obtain one entangled state over the total repeater distance, after optimizing nesting level. The black dotted line is for direct transmission. The green curve is for the DLCZ scheme. The blue dashed, dotted, and solid curves are for our scheme with $M=10$, $M=100$, and $M=1000$. For the DLCZ scheme the fidelity of the obtained state is $F=0.9$ whereas in the absence of multi-photon errors the fidelity of remote state in our scheme is $F=1$.}
  {\label{fig:RateCalc}
}
\end{figure} 

In order to facilitate comparisons with previous work, our numerical results employ similar assumptions, methods, and parameter values as those used in the review paper by Sangouard et al.\cite{sangouard_quantum_2011}, with one important additional factor to account for the disparate quantum channel phase sensitivity of the two schemes. The enormous reduction in phase sensitivity (by $10^7$) for two-photon interference over single-photon interference is critical for any realistic implementation, and we account for this with an additional source of inefficiency, $\eta_\phi=0.001$ for DLCZ. This value is quite reasonable, and has precedent as discussed in Ref. \citeonline{jiang_fast_2007}. Except for this parameter our assumptions are otherwise the same, and include number resolving detectors with detector efficiency $\eta_d=0.9$, memory readout efficiency $\eta_m=0.9$, speed of light in fiber $c=\SI[parse-numbers=false]{2\times10^{8}}{\meter\per\second}$, and a value of $p=0.01$ for the probability of generating a spin-wave and Stokes photon entangled state as in Equation \ref{ensemble_excite}. $M$, the number of multiplexed memory modes accessible to use, is varied between values of $M=10$ and $M=1000$. We optimize the nesting level $n$, thus the number of links, and limit the number of elementary segments to 16 corresponding to $n=4$. Finally, despite our above discussion of current state-of-the-art wavelength conversion technology, we follow prior convention and assume perfect conversion efficiency from near-infrared rubidium wavelengths ($780nm$) to telecom ($1550nm$), corresponding to fiber attenuation of \SI{0.2}{\decibel\per\kilo\meter}. A comparison of the times to obtain a remote entangled state for our scheme, DLCZ, and direct transmission is shown in Figure \ref{fig:RateCalc}.

As with all quantum communication rate analyses to date, Figure \ref{fig:RateCalc} shows that even with a multiplexed ensemble the time to generate useful entangled states over large distances is slow (nearly $10^2$ seconds) when compared with modern classical data rates.  However, the improvements from multiplexing are clear, and will also be generally applicable to faster and more sophisticated entanglement generation schemes. Such schemes may involve deterministic single photon sources, cavity-mediated quantum gates, or continuous-variable qubits.  In our opinion, efficient multiplexed interfaces will likely be a key ingredient to any future networked quantum systems.  

Based on the scheme described here, we envision creating a three-node quantum network, realizing a quantum repeater with only three multiplexed atom-cavity systems in remote locations (as opposed to the 8 required by the DLCZ protocol). These atom-cavity nodes require fewer resources, and are arguably simpler and more cost effective than competing ion-based platforms.  The possibility of demonstrating a quantum repeater \emph{in the near term} is an alluring and valuable scientific goal. There remain many basic scientific questions about how to proceed in building large scale entangled networks, not to mention what their ultimate purpose or value will be. But to address these questions it will be critical to build a functional quantum network to explore the application space. Our proposed scheme is a step forward on this journey.

\acknowledgments 
We thank Fredrik Fatemi for useful discussions. This work is supported by the US Army Research laboratory and the Quantum Science and Engineering Program from the US Office of the Secretary of Defense.

\bibliography{SPIEMultiplexedCavityRepeater} 

\begin{thebibliography}{10}

\bibitem{duan_long-distance_2001}
Duan, L.-M., Lukin, M.~D., Cirac, J.~I., and Zoller, P., ``Long-distance
  quantum communication with atomic ensembles and linear optics,'' {\em
  Nature}~{\bf 414},  413--418 (Nov. 2001).

\bibitem{sangouard_quantum_2011}
Sangouard, N., Simon, C., {de Riedmatten}, H., and Gisin, N., ``Quantum
  repeaters based on atomic ensembles and linear optics,'' {\em Reviews of
  Modern Physics}~{\bf 83},  33--80 (Mar. 2011).

\bibitem{blinov_observation_2004}
Blinov, B.~B., Moehring, D.~L., Duan, L.-M., and Monroe, C., ``Observation of
  entanglement between a single trapped atom and a single photon,'' {\em
  Nature}~{\bf 428},  153--157 (Mar. 2004).

\bibitem{reiserer_cavity-based_2015}
Reiserer, A. and Rempe, G., ``Cavity-based quantum networks with single atoms
  and optical photons,'' {\em Reviews of Modern Physics}~{\bf 87},  1379--1418
  (Dec. 2015).

\bibitem{bernien_heralded_2013}
Bernien, H., Hensen, B., Pfaff, W., Koolstra, G., Blok, M.~S., Robledo, L.,
  Taminiau, T.~H., Markham, M., Twitchen, D.~J., Childress, L., and Hanson, R.,
  ``Heralded entanglement between solid-state qubits separated by three
  metres,'' {\em Nature}~{\bf 497},  86--90 (May 2013).

\bibitem{usmani_mapping_2010}
Usmani, I., Afzelius, M., {de Riedmatten}, H., and Gisin, N., ``Mapping
  multiple photonic qubits into and out of one solid-state atomic ensemble,''
  {\em Nature Communications}~{\bf 1},  12 (Apr. 2010).

\bibitem{zhou_quantum_2015}
Zhou, Z.-Q., Hua, Y.-L., Liu, X., Chen, G., Xu, J.-S., Han, Y.-J., Li, C.-F.,
  and Guo, G.-C., ``Quantum {{Storage}} of {{Three}}-{{Dimensional
  Orbital}}-{{Angular}}-{{Momentum Entanglement}} in a {{Crystal}},'' {\em
  Physical Review Letters}~{\bf 115},  070502 (Aug. 2015).

\bibitem{gorshkov_universal_2007}
Gorshkov, A.~V., Andr\'e, A., Fleischhauer, M., S\o{}rensen, A.~S., and Lukin,
  M.~D., ``Universal {{Approach}} to {{Optimal Photon Storage}} in {{Atomic
  Media}},'' {\em Physical Review Letters}~{\bf 98} (Mar. 2007).

\bibitem{collins_multiplexed_2007}
Collins, O.~A., Jenkins, S.~D., Kuzmich, A., and Kennedy, T. A.~B.,
  ``Multiplexed {{Memory}}-{{Insensitive Quantum Repeaters}},'' {\em Physical
  Review Letters}~{\bf 98},  060502 (Feb. 2007).

\bibitem{dudin_light_2013}
Dudin, Y.~O., Li, L., and Kuzmich, A., ``Light storage on the time scale of a
  minute,'' {\em Physical Review A}~{\bf 87} (Mar. 2013).

\bibitem{olmschenk_quantum_2009}
Olmschenk, S., Matsukevich, D.~N., Maunz, P., Hayes, D., Duan, L.-M., and
  Monroe, C., ``Quantum teleportation between distant matter qubits,'' {\em
  Science}~{\bf 323}(5913),  486--489 (2009).

\bibitem{ritter_elementary_2012}
Ritter, S., N\"olleke, C., Hahn, C., Reiserer, A., Neuzner, A., Uphoff, M.,
  M\"ucke, M., Figueroa, E., Bochmann, J., and Rempe, G., ``An elementary
  quantum network of single atoms in optical cavities,'' {\em Nature}~{\bf
  484},  195--200 (Apr. 2012).

\bibitem{nolleke_efficient_2013}
N\"olleke, C., Neuzner, A., Reiserer, A., Hahn, C., Rempe, G., and Ritter, S.,
  ``Efficient {{Teleportation Between Remote Single}}-{{Atom Quantum
  Memories}},'' {\em Physical Review Letters}~{\bf 110} (Apr. 2013).

\bibitem{bao_efficient_2012}
Bao, X.-H., Reingruber, A., Dietrich, P., Rui, J., D\"uck, A., Strassel, T.,
  Li, L., Liu, N.-L., Zhao, B., and Pan, J.-W., ``Efficient and long-lived
  quantum memory with cold atoms inside a ring cavity,'' {\em Nature
  Physics}~{\bf 8},  517--521 (July 2012).

\bibitem{laurat_towards_2007}
Laurat, J., Chou, C.-w., Deng, H., Choi, K.~S., Felinto, D., de~Riedmatten, H.,
  and Kimble, H.~J., ``Towards experimental entanglement connection with atomic
  ensembles in the single excitation regime,'' {\em New Journal of
  Physics}~{\bf 9},  207--207 (June 2007).

\bibitem{choi_mapping_2008}
Choi, K.~S., Deng, H., Laurat, J., and Kimble, H.~J., ``Mapping photonic
  entanglement into and out of a quantum memory,'' {\em Nature}~{\bf 452},  67
  (Mar. 2008).

\bibitem{chrapkiewicz_high-capacity_2017}
Chrapkiewicz, R., D\k{a}browski, M., and Wasilewski, W., ``High-{{Capacity
  Angularly Multiplexed Holographic Memory Operating}} at the
  {{Single}}-{{Photon Level}},'' {\em Physical Review Letters}~{\bf 118},
  063603 (Feb. 2017).

\bibitem{pu_experimental_2017}
Pu, Y.-F., Jiang, N., Chang, W., Yang, H.-X., Li, C., and Duan, L.-M.,
  ``Experimental realization of a multiplexed quantum memory with 225
  individually accessible memory cells,'' {\em Nature Communications}~{\bf 8},
  ncomms15359 (May 2017).

\bibitem{tian_spatial_2017}
Tian, L., Xu, Z., Chen, L., Ge, W., Yuan, H., Wen, Y., Wang, S., Li, S., and
  Wang, H., ``Spatial {{Multiplexing}} of {{Atom}}-{{Photon Entanglement
  Sources}} using {{Feedforward Control}} and {{Switching Networks}},'' {\em
  Physical Review Letters}~{\bf 119},  130505 (Sept. 2017).

\bibitem{yang_multiplexed_2018}
Yang, T.-S., Zhou, Z.-Q., Hua, Y.-L., Liu, X., Li, Z.-F., Li, P.-Y., Ma, Y.,
  Liu, C., Liang, P.-J., Li, X., Xiao, Y.-X., Hu, J., Li, C.-F., and Guo,
  G.-C., ``Multiplexed storage and real-time manipulation based on a multiple
  degree-of-freedom quantum memory,'' {\em Nature Communications}~{\bf 9},
  3407 (Aug. 2018).

\bibitem{grodecka-grad_high-capacity_2012}
{Grodecka-Grad}, A., Zeuthen, E., and S\o{}rensen, A.~S., ``High-{{Capacity
  Spatial Multimode Quantum Memories Based}} on {{Atomic Ensembles}},'' {\em
  Physical Review Letters}~{\bf 109},  133601 (Sept. 2012).

\bibitem{parniak_wavevector_2017}
Parniak, M., D\k{a}browski, M., Mazelanik, M., Leszczy\'nski, A., Lipka, M.,
  and Wasilewski, W., ``Wavevector multiplexed atomic quantum memory via
  spatially-resolved single-photon detection,'' {\em Nature
  Communications}~{\bf 8},  2140 (Dec. 2017).

\bibitem{simon_interfacing_2007}
Simon, J., Tanji, H., Thompson, J.~K., and Vuleti\'c, V., ``Interfacing
  {{Collective Atomic Excitations}} and {{Single Photons}},'' {\em Physical
  Review Letters}~{\bf 98} (May 2007).

\bibitem{bimbard_homodyne_2014}
Bimbard, E., Boddeda, R., Vitrant, N., Grankin, A., Parigi, V., Stanojevic, J.,
  Ourjoumtsev, A., and Grangier, P., ``Homodyne {{Tomography}} of a {{Single
  Photon Retrieved}} on {{Demand}} from a {{Cavity}}-{{Enhanced Cold Atom
  Memory}},'' {\em Physical Review Letters}~{\bf 112} (Jan. 2014).

\bibitem{cox_increased_2018}
Cox, K.~C., Meyer, D.~H., Schine, N.~A., Fatemi, F.~K., and Kunz, P.~D.,
  ``Increased atom-cavity coupling and stability using a parabolic ring
  cavity,'' {\em Journal of Physics B: Atomic, Molecular and Optical
  Physics}~{\bf 51},  195002 (Oct. 2018).

\bibitem{thompson_observation_1992}
Thompson, R.~J., Rempe, G., and Kimble, H.~J., ``Observation of normal-mode
  splitting for an atom in an optical cavity,'' {\em Physical Review
  Letters}~{\bf 68},  1132--1135 (Feb. 1992).

\bibitem{hunger_fiber_2010}
Hunger, D., Steinmetz, T., Colombe, Y., Deutsch, C., H\"ansch, T.~W., and
  Reichel, J., ``A fiber {{Fabry}}\textendash{{Perot}} cavity with high
  finesse,'' {\em New Journal of Physics}~{\bf 12}(6),  065038 (2010).

\bibitem{lan_multiplexed_2009}
Lan, S.-Y., Radnaev, A.~G., Collins, O.~A., Matsukevich, D.~N., Kennedy, T.
  a.~B., and Kuzmich, A., ``A {{Multiplexed Quantum Memory}},'' {\em Optics
  Express}~{\bf 17},  13639--13645 (Aug. 2009).

\bibitem{pu_experimental_2018}
Pu, Y., Wu, Y., Jiang, N., Chang, W., Li, C., Zhang, S., and Duan, L.,
  ``Experimental entanglement of 25 individually accessible atomic quantum
  interfaces,'' {\em Science Advances}~{\bf 4},  eaar3931 (Apr. 2018).

\bibitem{chen_demonstration_2007}
Chen, S., Chen, Y.-A., Zhao, B., Yuan, Z.-S., Schmiedmayer, J., and Pan, J.-W.,
  ``Demonstration of a {{Stable Atom}}-{{Photon Entanglement Source}} for
  {{Quantum Repeaters}},'' {\em Physical Review Letters}~{\bf 99},  180505
  (Nov. 2007).

\bibitem{radnaev_quantum_2010}
Radnaev, A.~G., Dudin, Y.~O., Zhao, R., Jen, H.~H., Jenkins, S.~D., Kuzmich,
  A., and Kennedy, T. a.~B., ``A quantum memory with telecom-wavelength
  conversion,'' {\em Nature Physics}~{\bf 6}(11),  894--899 (print November
  2010).

\bibitem{farrera_nonclassical_2016}
Farrera, P., Maring, N., Albrecht, B., Heinze, G., and de~Riedmatten, H.,
  ``Nonclassical correlations between a {{C}}-band telecom photon and a stored
  spin-wave,'' {\em Optica}~{\bf 3},  1019--1024 (Sept. 2016).

\bibitem{chen_fault-tolerant_2007}
Chen, Z.-B., Zhao, B., Chen, Y.-A., Schmiedmayer, J., and Pan, J.-W.,
  ``Fault-tolerant quantum repeater with atomic ensembles and linear optics,''
  {\em Physical Review A}~{\bf 76} (Aug. 2007).

\bibitem{lutkenhaus_bell_1999}
L\"utkenhaus, N., Calsamiglia, J., and Suominen, K.-A., ``Bell measurements for
  teleportation,'' {\em Physical Review A}~{\bf 59},  3295--3300 (May 1999).

\bibitem{jiang_fast_2007}
Jiang, L., Taylor, J.~M., and Lukin, M.~D., ``Fast and robust approach to
  long-distance quantum communication with atomic ensembles,'' {\em Physical
  Review A}~{\bf 76} (July 2007).

\end{thebibliography}
\bibliographystyle{spiebib} 

\end{document}